# Tuning Multiple Motor Travel Via Single Motor Velocity

Jing Xu[1,2,*], Zhanyong Shu[1], Stephen J. King[3], and Steven P. Gross[1,*]

[1]Developmental and Cell Biology, School of Biological Sciences, University of California, Irvine, CA 92697, USA
[2]School of Natural Sciences, University of California, Merced, CA 95343, USA
[3]Burnett School of Biomedical Sciences, University of Central Florida, Orlando, FL 32827, USA

*Corresponding Authors: Steven Gross, sgross@uci.edu; Jing Xu, jing.xu@ucmerced.edu

**Running Title:** Velocity Control of Travel Distance

## Abstract

Microtubule-based molecular motors often work in small groups to transport cargos in cells. A key question in understanding transport (and its regulation *in vivo*) is to identify the sensitivity of multiple-motor-based motion to various single molecule properties. Whereas both single-motor travel distance and microtubule binding rate have been demonstrated to contribute to cargo travel, the role of single-motor velocity is yet to be explored. Here, we recast a previous theoretical study, and make explicit a potential contribution of velocity to cargo travel. We test this possibility experimentally, and demonstrate a strong negative correlation between single-motor velocity and cargo travel for transport driven by two motors. Our study thus discovers a previously unappreciated role of single-motor velocity in regulating multiple-motor transport.

## Introduction

Active intracellular transport is crucial for cell function, and defects are linked to diseases including neurodegeneration. Frequently, *in vivo* cargos are moved by multiple microtubule-based molecular motors (*1-6*). Single molecule biophysical studies have revealed a great deal about the function of individual motors *in vitro*, but it is still a challenge to use single molecule properties to explain group function. Intuitively, when only one type of motor is present, multiple motors working together ought to improve transport relative to the single motor. While this is in general supported by theoretical (*7-10*) and experimental (*11-15*) studies, exactly how multiple motors work together in groups still remains an area of active research (*14, 16-19*). In particular, a key open question is how to **regulate** multiple-motor transport. That is, which single-molecule properties are crucial for determining the motors' group functions, and how altering these properties might impact their group functions.

One way to better understand the functions of groups of motors—and thus which single-molecule properties may be essential—is to develop a validated theoretical description of multiple-motor based transport. There have been two competing approaches, either using a mean-field approach assuming motors share load equally (*7*), or by taking a Monte-Carlo stochastic approach that does not make such assumptions (*8, 10*). These two formalisms make different predictions for travel under significant load (i.e. for loads comparable to single motor stall force), and under such extreme conditions the Monte-Carlo approach seems to better describe some aspects of *in vitro* motor function (*8, 10*). However, for travel under no load or low load, the two formalisms converge, and in this case, the mean-field models are preferable because their analytic results allow greater insight into the effects of alteration of single-molecule properties. In the current study, we thus focus on travel under no load and re-interpret a previous analytic description/prediction (*7*). We then experimentally test the resulting prediction for the two-motor system. Our study uncovers a surprising link between motor velocity and transport, and demonstrates that single-motor **velocity** can be used to significantly tune multiple-motor transport.

## Results

### *Recasting the Two-Motor Travel Prediction in Crucial Single Motor Properties*

The analytic description of multiple-motor transport developed using mean field theory (*7*) is quite general, but sufficiently complicated that its implications are not immediately obvious. Here we first summarize the analytic description for the two motor case, and then rewrite it as a function of measurable and independent single-molecule parameters to better understand and compare with experiments.

When two motors are available for transport, the cargo may be linked via one or both motors to the microtubule at any instance during its travel (Fig. 1). How far this cargo can travel depends crucially on the transition rates of the cargo between the one-motor and both-motor bound states. The greater the transition rate into the two-motor bound state, the further the cargo will travel in excess of what would occur due to a single motor alone. For two identical, non-interacting motors, these transition rates are simplified to depend only on individual motors' binding and unbinding rates for the microtubule (Fig. 1). In turn, the average travel for a cargo carried by two non-interacting motors (*D*), as developed by mean field theory (*7*), takes the analytic expression,

$$D = \frac{v}{\varepsilon}(1 + \frac{v_2}{v} \cdot \frac{\pi_{ad}}{2\varepsilon}) \qquad \text{Eqn. (1)}$$

Here, $\varepsilon$ and $\pi_{ad}$ are the single-motor unbinding and binding rates for microtubules (s$^{-1}$), respectively. $v$ is the single-motor velocity, and $v_2$ is the velocity of cargo when driven simultaneously by both available motors.

For processive motors that do not interact with each other, the ensemble velocity is constant and identical to the single motor velocity, that is, $v_2 = v$. With this simplification, the average two-motor travel in Eqn. (1) now depends only on single motor properties, and can be written as

$$D = \frac{v}{\varepsilon}(1 + \frac{\pi_{ad}}{2\varepsilon})$$  Eqn. (2).

Eqn. (2) provides a simplified theoretical framework to approach two-motor travel. Intriguingly, it suggests a previously unexplored role of single motor velocity ($v$) for tuning two-motor travel ($D$). However, since the single-motor unbinding rate ($\varepsilon$) can depend non-trivially on the motor's velocity ($v$) (see below), Eqn. (2) does not yet lead to predictions easily tested by experiments.

To further simplify the mean field description in Eqn. (2) into independent parameters, we assume a constant probability of unbinding per step for the motor. In this scenario, the motor associates tightly to the microtubule when it is paused between steps. This is a reasonable assumption for kinesin-based transport (*15, 20*) (also verified in the current study). The motor's unbinding rate per unit time ($\varepsilon$) is then a product of its unbinding rate per step ($\Delta d/d$) and its stepping rate ($v/\Delta d$). Here $\Delta d$ denotes the motor's step size, and $d$ denotes the average distance a single motor can travel before becoming unbound from the microtubule (single-motor travel distance). We thus arrive at a simple relationship, $\varepsilon = v/d$, and recast Eqn. (2) as

$$D = d(1 + \frac{\pi_{ad}d}{2v})$$  Eqn. (3).

Note that Eqn. (3) makes clear predictions on how the two-motor travel distance ($D$) relates to the single-motor travel ($d$): the average two-motor travel is $(1 + \frac{\pi_{ad}d}{2v})$–fold greater than single-motor travel. Since the velocity term ($v$) appears in the denominator and does not affect the motor's travel distance ($d$) or microtubule binding rate ($\pi_{ad}$), we predict increasing two-motor travel distance with reducing single motor velocity. This prediction is perhaps counterintuitive, given that the cargo is driven by the same set of motors under the same microtubule conditions. In the current study, we experimentally test this theoretical prediction, considering velocity as a potential tuning parameter for multiple-motor travel.

### *A New Assay to Experimentally Measure Two-Kinesin Transport*

We employed conventional kinesin as an experimental system to test the predictions of Eqn. (3). To cleanly compare theory and experiment, we need to achieve the required geometry for both single-kinesin and two-kinesin transport.

Single-kinesin geometry is well developed and studied. In particular, antibodies have been used to specifically recruit tagged motor proteins onto beads (*21-24*) (illustrated in Fig. 2A). In these experiments, beads are first functionalized with penta-his antibodies, surface blocked (to eliminate non-specific motor adsorption, or motor-independent bead-microtubule interaction), and then incubated with C-term his-tagged kinesins. The antibody presence on beads is kept in excess of available motors, to minimize the probability of two motors being recruited by the same antibody. To reach the single motor regime, motor presence on beads is titrated such that the fraction of beads capable of binding to microtubules is limited to below 50% (*24-25*). We utilize this method/geometry to evaluate one-motor transport (Fig. 2A(i), 2B).

The geometry required for two-motor transport remains an area of intense experimental focus (*26-27*). In our case, the cargo should be driven by two and only two kinesin motors, attached in close proximity to each other, so that both can reach the microtubule simultaneously. Here we take advantage of the two equivalent antigen binding sites present on the same antibody, and reverse the available motor-to-antibody ratio used for the current single motor assays (Fig. 2A(ii)). We note that this approach is similar to a previous study using reduced antibody: bead ratio to recruit the tetrameric Eg5 motor (*28*). Here, we specifically titrate the antibody presence on individual beads to the single antibody range (below 50% binding (*24-25*), Fig. 2B), so that only a single antibody can contribute to recruiting motors, and thus, no more than two motors can simultaneously contribute to transport. To increase the occurrence of the two-motor configuration, we adjust the motor concentration to be in significant excess of the amount antibodies present. We verify that all motors measured are recruited via the antibody: in the absence of antibody, we observe no bead binding to microtubules, despite excess kinesin in solution.

We then employed an optical trap to measure the subset of beads with two active motors recruited by the single antibody. Using the additive property of motor force production (*4, 11-12, 17, 29-30*), we experimentally adjusted the optical trap to be strong enough to trap a bead driven by one, but not two kinesins together (Fig. 3). For beads to successfully escape this one-motor trap, it needs to persistently exert force greater than 5pN (single-kinesin force (*17, 23, 30-31*)) and continuously move away from the trap center. For these beads, we turn off the trap as soon as possible (indicated by green arrows, Fig 3) to enable motility measurements of beads carried by two kinesin motors in the absence of external load. As a consequence of this force selection, our minimal measurable travel distance for the two-motor case is ~0.6μm.

### *Single-Kinesin Processivity Does Not Correlate with Velocity*

We started with control experiments, and characterized the effect of different velocities on single-motor transport (Fig. 4A). To tune velocity for wild-type kinesin motors, we simply altered the ATP concentrations in otherwise identical assays. At the single-molecule level, we verified that decreased ATP concentration resulted in decreased average bead velocity (Fig 4A). The measured velocity dependence on ATP concentration is well characterized by Michaelis–Menten kinetics ($K_m = 44\mu M$, and $v_{max} = 0.9\mu m/s$).

At saturating ATP, the measured single kinesin travel distance is in good agreement with previous reports for the same construct (*30, 32*). At the two lower ATP concentrations tested, the average travel of individual kinesin remained constant (1.67µm average) and was unaffected by the change in velocity (Fig. 4A). This observed constant single-kinesin travel distance satisfies our theory assumption of a velocity-dependent unbinding rate $\varepsilon = v/d$ (Fig. S1). It is also consistent with previous findings using either ATP concentration (*20*) or crowding effects on the microtubule (*15*) to reduce kinesin's velocity. Thus, any effect observed for two-motor travel must result from changes in group function as opposed to single-molecule effects.

### *Average Cargo Velocity Not Affected by Motor Copy*

The average velocity of beads driven by two-motors was similarly affected by ATP levels, with lower ATP levels again resulting in slower velocity (see Fig. 3 for example traces, and Fig. 4B, left for histograms). We observe no significant difference in the average velocities for beads carried by one- *vs.* two-kinesins. This is as expected for the highly processive motors (such as kinesin) traveling under no load (*6, 11-12, 14, 33*), and satisfies our theoretical simplification that $v = v_2$.

### *Two-Kinesin Travel Negatively Correlates with Velocity*

Remarkably, beads carried by two kinesins traveled further with reducing motor velocity (Fig. 4B). Under otherwise *identical* two-motor conditions (Fig. 2A(ii)), reductions in the ATP concentration significantly increased the population of beads walking out of our field of view from zero to more than 40% of the measured population (Fig. 4B, right). Since the single kinesin travel is not affected by velocity, the current study directly demonstrates velocity as a tuning parameter for two-motor travel, consistent with that predicted by Eqn. (3).

At saturating ATP, the average two-kinesin travel distance (*D*=2.95 µm) is 1.77-fold that of the single kinesin travel (*d*=1.67µm). This modest increase is in excellent agreement with a previous study (*14*) coupling two kinesins together via a short DNA linkage. There, a 1.73-fold travel increase was observed under similar saturating ATP conditions, but any potential link between single-motor velocity and ensemble travel distance was not considered. None-the-less, the agreement between these two independent studies (using distinct approaches and geometries) indicates that there is negligible three- or more motor contamination in our assay.

### *Quantitative Comparison between Theory and Experiment*

Could the increase in two-motor travel arise from altered motor-motor interaction at lower velocities? Theory development (Eqn. 1-2) assumed that the motors do not interact with each other. In experiments, however, there is likely some interaction. For example, the mechanical property of recruitment linkage can alter individual motor's binding rate for microtubules (*16*). Such potential interaction is of interest since, in principle, it can alter the motor's binding and unbinding rate in ensemble. The exact detail of motor-motor interaction in an ensemble is best probed by travel measurements under significant external load (such as (*8, 17-19*)), and is not the focus of the current study. However, we can use quantitative comparison with theory to address whether or not our data ***requires*** the motor-motor interaction to vary with velocity.

For simplicity, we assume that any potential motor-motor interaction is constant (velocity-independent), and that the interaction alters the binding rate ($\pi_{ad}$) but not travel distance ($d$) of individual motors in ensemble. We treat this motor's binding rate in an ensemble as an unknown parameter, that we determine/constrain using measurements at $v_{max} = 0.9\mu m/s$. This approach allows us to utilize the measured single motor travel distance, and determine the necessary term $\pi_{ad}d$ in Eqn. (3). We then use Eqn. (3) to make constrained predictions (with no tunable/free parameters) for what should occur as the motor slows down from $v_{max}$ to lower velocities. Since theory assumes no interactions between the motors, and we account for any unchanging interactions by constraining the binding rate, any quantitative disagreement between theory and measurements would indicate the likelihood of a velocity-dependent interaction.

In addition to the average cargo travel discussed thus far (Eqn. 1-3), the complete theory developed using a mean-field theory (*7*) also predicts a bi-phasic distribution for two-motor travel. That is, we expect the travel distribution to be described by the sum of two exponential decays. Since different combinations of two exponentials can give rise to the same average travel, we now use the measured two-motor travel distribution instead of its average value to constrain the single-motor binding rate ($\pi_{ad}$). For non-interacting motors, the predicted two-motor travel distribution ($\psi$) (*7*) reduces to,

$$\psi(x) = c_a\, e^{-x/\lambda_a} + c_b\, e^{-x/\lambda_b} \qquad \text{Eqn. (4)}$$

where $c_a, c_b, \lambda_a$, and $\lambda_b$ are functions of single-motor binding rate ($\pi_{ad}$), travel ($d$), and velocity ($v$) (see Methods for explicit forms).

The measured travel distribution at $v_{max} = 0.9\mu m/s$ is well contained within our field of view (Fig 4B, "1mM ATP"). Using Eqn. (4) to model the measured two-kinesin travel distribution at $v_{max} = 0.9\mu m/s$ (Fig. 4B, "1mM ATP"), we obtain a least-$\chi^2$ fitted value of $\pi_{ad}$= 0.71/s (Fig. 5A). The corresponding best-fit decay lengths $\lambda_a$ and $\lambda_b$ are 3.16 and 0.44µm, respectively (scatter, Fig. 5B). This shorter decay length ($\lambda_b$=0.44µm) is significantly below our lower measurement threshold (0.6µm). The resulting two-motor distribution (Fig. 5C, "0.9µm/s") recaptures the experimental measurement (scatter, Fig. 5C), and is largely described by a single exponential above our 0.6µm measurement threshold.

Using the now-constrained single motor binding rate $\pi_{ad}$= 0.71/s, we numerically evaluated the two decay lengths in Eqn. (4) ($\lambda_a$ and $\lambda_b$, explicit form in Methods), and found them to diverge with reducing velocities (Fig. 5B). The slower the velocity, the shorter we expect $\lambda_b$ to be, and the longer we expect $\lambda_a$ to be. At the lowest measured velocity (0.16μm/s), the long decay length exceeds our field of view, with $\lambda_a$=8.53μm, and the short decay length is less than a third of our minimum measurement threshold at $\lambda_b$=0.16μm (Fig. 5B-C). For experiments, this extremely biphasic distribution translates to an expected low number of counts within our precision measurement range (Fig. 5C), and is consistent with the limited counts we observed in each measurement bin (Fig. 4B). While specific quantitation of travel lengths was not possible, all travels exceeding our field of view were recorded (the beads were seen to walk out of the camera view), and make for a striking contrast between measurements at $v_{max} = 0.9$μm/s, and at the two lower velocities (0.26 and 0.16μm/s). We thus evaluated the fraction of beads traveling beyond our field of view (>7.6μm) *vs.* the total measurable population (>0.6μm). This strategy allows us to utilize measurement counts in a well defined quantitative manner, with reduced error (by accumulating measurement counts across bins). Similarly, by integrating the predicted travel distributions over the same range (>0.6μm, and >7.6μm), we obtain the predicted population of long travel for the two lower velocities. We observe excellent quantitative agreement between measurements and theory (Fig. 5D). Thus, the observed increase in two-motor travel distance is unlikely due to altered motor-motor interaction at reduced velocities.

**Discussion**

We present here the first experimental demonstration that single-motor velocity is an important control for multiple-motor transport. At saturating ATP, we observe a modest increase in cargo travel due to the second motor (1.77-fold, Fig. 4), similar to a previous report (*14*). However, when we reduce single motor velocity, the average two-motor travel distance is significantly increased (Fig. 4). This dramatic increase in two-motor travel is in excellent quantitative agreement with current theoretical predictions (Eqn. 3-4, Fig. 5D), and does not appear to require potential motor-motor interaction to vary with velocity. Thus, for multiple-motor driven transport, perhaps surprisingly, one can increase the average cargo travel distance simply by decreasing single-motor velocity.

The negative correlation between single-motor velocity and multiple-motor travel may be understood intuitively. For the second available motor to contribute to motion, the cargo must be linked by both motors to the microtubule for some fraction of its travel. The larger this fraction, the more contribution from the second motor, and the further the overall transport will be. This fraction depends on the relative rates of the cargo transitioning into, and out of the two-motor-bound state ($\pi_{ad}$ and $2\varepsilon$, respectively, Fig. 1). In the simple case considered in this study, the unbinding rate ($\varepsilon$) is inversely tuned by velocity (Fig. 4A & S1). Thus by decreasing velocity, we reduce the rate of cargo transitioning out of the two-motor bound state ($2\varepsilon$, Fig. 1), thereby enhancing the contribution from the additional motor.

Note that a velocity-independent unbinding rate is inconsistent with the measured single-kinesin travel distance (Fig. S1). Our study indicates that the kinesin's unbinding rate is dominated by its probability of to fall off the microtubule during stepping, and that any contribution from its waiting state (between steps) is negligible. Since the motor's stepping rate is directly proportional to its velocity, the unbinding rate ($\varepsilon$) of the kinesin motor must be velocity-dependent, the transition rate out of the two- (and more-) kinesin bound state must also be tunable by the velocity.

The impact of velocity on multiple-motor travel does not need to be limited to kinesin-based transport. For example, the single motor travel of some mamalian myosins is found to *increase* with reducing velocity (*34-35*). Our study would suggest that multiple-motor travel for these myosins would become even more enhanced with decreasing velocity than for kinesin.

Although for experimental ease we use alteration of ATP levels to change velocity, in principle, motor velocity *in vivo* could be regulated in multiple ways. Changes in a motor's mechano-chemical cycle (achieved e.g. by motor phosphorylation, or the presence of cofactors) could alter its velocity when functioning at the same ATP level, and indeed such regulation is established to occur in other contexts (reviewed in (*36*)). Further, several studies using mutations to modulate the mechano-chemical cycle of motors have introduced significant velocity alterations *in vitro* (*37-39*). Our study suggests that velocity tuning may be a useful strategy to overcome single motor limitations, and that modest velocity reduction has the potential to significantly benefit multiple-motor travel. For example, limiting the average velocity to ~0.3μm/s would in principle be sufficient to negate single motor processivity defect for homozygous mutant dynein identified in our previous murine dynein study (*6*), and recover wild-type lysosomal travel distance in axons. Whether and how biology might take advantage of this mode of regulation remains to be explored.

Little is known about the biological relevance of velocity diversity in molecular motors. Our study suggests an interesting trade-off: at the cost of taking longer, by slowing down, the same group of motors can on average transport a cargo further. This mode of potential regulation may be relevant for slow *vs.* fast transport in axons, and future high resolution studies may identify specific classes of axonal cargos with sustained travel at reduced instantaneous velocities. An additional implication for transport regulation *in vivo* concerns cargos moved by a mixture of motors of the same directionality but drastically different velocities (*40*). A recent study demonstrates that the relative density of slow *vs.* fast motors provides regulation to select for transport dominated by one or the other motor (*41*). Our study introduces an enticing possibility: by tuning the relative number of motors in each of the sets, the cell can control velocity, and can thus tune the average transport distance for the particular cargo in question. While there are indeed many different cargo velocities observed *in vivo* (see example (*42*)), it remains to be determined the extent to which mean travel is regulated via control of single-motor velocity.

**Materials and Methods**

## Protein purification

Expression construct of the recombinant kinesin K560 (*43*), pET17b-K560-6xhis, was constructed using plasmid pET17b_k560_GFP_his (Addgene) as template. Nde1 and Xho1 sites were used as restriction sites, and the DNA oligo sequences used for PCR are listed as 5'-GATATACATATGGCGGACCTGGCCGAG, and 5'-GCGGGGCTCGAGTTAATGGTGGTGGTGATGATGTTTTAGTAAAGATGCCATCATC. Vector pET17 was obtained from Novagen. Bacteria strain XL1-blue competent cells (Stratagene) was used for the transformation of pET17b-k560-6xhis. Sequence verified constructs was transformed into *E. coli* strain BL21 DE3 Rosetta for protein expression. Cells was selected by ampicilin and chloramphenicol, induced for expression at 18 °C, lysed by sonication in ice water, and clarified by ultra centrifugation at 30,000 RCF. Protein was purified by Ni-NTA affinity column by first washing with 50mM phosphate buffer pH 8.0, 300mM NaCl and 75mM Imidazole, and eluted out with 50 mM phosphate buffer pH 8.0, 300mM NaCl and 250mM Imidazole.

Bovine tubulin was purified over a phosphocellulose column as described (*44*). Recombinant Casein Kinase 2 was purchased from New England Biolabs.

## Optical Trapping Assay

Bead assay and data analysis were carried out as previously described (*6, 12, 29*), except the motor/bead recruitment approach used. In the current study, kinesin motors were specifically recruited via monoclonal antibody against C-terminal his-tag of the recombinant K560. Antibody presence on bead was further tuned to achieve either the single-motor or two-motor geometry (Fig. 2). To ensure that the majority of the kinesin motors in our assay are active, we preincubate K560 motors with casein kinase 2 (3:1 kinase:motor ratio) on ice for 1.5 hr prior to incubating with anti-His beads in all experiments. Such pre-incubation with casein kinase 2 is reported to enhance K560 activity without altering its single-motor functions (*30*).

To measure bead binding fraction, we use a single beam optical trap to position individual beads near microtubules (preassembled in flow cells) for 30 seconds. A binding event is scored if the bead binds to the microtubule, and processes away from the trap center within the 30 seconds wait time. This waiting time is sufficiently long compared to the bead's rotation rate while in trap and individual motor binding rate (*25*). The measurement of a binding event is insensitive to any increase in the bead's microtubule binding rate resulting from two motors are recruited by the same antibody.

For both one-, and two-kinesin transport, the average cargo velocity was determined by Gaussian fitting to the measured velocity distribution, and the average travel determined by fitting a single exponential to the measured travel distributions where appropriate. For travel distributions with significant population escaping our field of view (Fig. 4B, "10μM ATP" and "20μM ATP"), we evaluated the minimum algorithmic mean and associated error.

## Explicit Form for Predicted Two-Motor Travel Distribution (7)

The analytic descriptions developed by the mean field approach describes the travel distribution driven by two motors as the sum of two exponential decays

$$\psi(x) = c_a e^{-x/\lambda_a} + c_b e^{-x/\lambda_b}.$$

For non-interacting motors, $c_a$, $c_b$, $\lambda_a$, and $\lambda_b$ can be rewritten as functions of single-motor binding rate ($\pi_{ad}$), travel distance ($d$), and velocity ($v$), using the same substitutions described in the text to derive Eqn. (3). Specifically,

$$c_a = \frac{\varepsilon}{2v}\left(1 - \frac{\pi_{ad}-\varepsilon}{R}\right) = \frac{1}{2d}\left(1 - \frac{\pi_{ad}d-v}{Rd}\right),$$
$$c_b = \frac{\varepsilon}{2v}\left(1 + \frac{\pi_{ad}-\varepsilon}{R}\right) = \frac{1}{2d}\left(1 + \frac{\pi_{ad}d-v}{Rd}\right),$$
$$\lambda_a = \frac{2v}{(3\varepsilon+\pi_{ad}-R)} = \frac{2vd}{(3v+\pi_{ad}d-Rd)},$$
$$\lambda_b = \frac{2v}{(3\varepsilon+\pi_{ad}+R)} = \frac{2vd}{(3v+\pi_{ad}d+Rd)},$$

and $R \equiv \sqrt{\varepsilon^2+\pi_{ad}^2 + 6\varepsilon\pi_{ad}} = \sqrt{(v/d)^2+\pi_{ad}^2 + 6v\pi_{ad}/d}$.

Least-$\chi^2$ fitting of the predicted (above, Eqn. 4) *vs.* measured (1mM ATP) two-motor travel distributions (Fig. 4B) were carried out using a custom-routine in Matlab, using $d$=1.67μm, $v$=0.9μm/s. The resulting best-fit $\pi_{ad}$ value allowed us to constrain the rate of individual motor binding to microtubules, and to evaluate the predicted two-motor travel distribution (Eqn. 4) at lower velocities.

## Acknowledgements

We thank Dr. Michael E. Fisher and Dr. Michael R. Diehl for helpful discussions. This work was supported by NIH grant RO1GM070676 to SPG, NIH grant NS048501 to SJK, and AHA grant 825278F to JX.

**Figure Legends**

**Figure 1.** Schematic illustration of two-motor transport. The cargo travel begins when at least one of two available motors become bound to the microtubule, and ends when neither motor is bound to the microtubule. During travel, the cargo can be linked to the microtubule via one motor, or both motors simultaneously. For two identical, non-interacting motors, the rates of transition between zero-, one-, and two-motor bound configurations were previously developed (*7*) and are indicated.

**Figure 2.** Experimental design for one-, and two-kinesin transport. (**A**) Illustrations of (**i**) one- and (**ii**) two-kinesin transport. Both utilize a monoclonal antibody (Y, against the C-term his-tag of recombinant K560) for specific motor recruitment. Both utilize casein (5.55mg/ml) to block antibody-independent, non-specific binding of motors onto beads. For single-kinesin transport (**i**) (*21-24*), the antibody presence on the bead is effectively saturated, to increase the density of motor binding sites; the motor presence was titrated to the single-motor range. For two-kinesin transport (**ii**), antibody presence per bead was titrated to be one antibody per bead, and excess kinesin in solution was used to increase the probability that two motors occupy both antigen binding sites of a single antibody. (**B**) The effect of motor or antibody presence on beads on the fraction of beads capable of binding to microtubules. Starting with excess antibody and excess kinesin per bead (such that all beads bound to microtubules), we significantly reduced the bead binding fraction either by reducing motor concentration (grey bars) or reducing antibody concentration (cyan bars). Thus any observed bead motion along microtubules must be mediated by the antibody-motor complex.

**Figure 3.** Representative position *vs.* time traces for a bead carried by two kinesin motors, measured at three distinct ATP concentrations. In the single-antibody range, each bead can recruit no more than two kinesins. We further utilize an optical trap to further select for beads carried by two and only two motors. We measured a ~5pN force production for the single kinesin motor, consistent with previous reports for the same construct (*17, 23, 30-31*). We thus adjust the optical trap (~2.5pN/100nm) to be strong enough to trap one kinesin motor (~5pN), but not two (~10pN). Traces within the highlighted grey region correspond to events that do not escape the optical trap. This includes one-motor stalls (red arrows), and two-motors attempts (blue arrows). Green arrows indicate where/when trap was turned off, after observing persistent motion beyond one motor stall, to enable the study of bead travel driven by two active motors under no load.

**Figure 4**. Velocity and travel distributions of beads carried by (**A**) one-, and (**B**) two-kinesins, measured at three ATP concentrations. Histograms are not normalized; Y-axis represents raw experimental counts. Average velocities and travel distances are shown in mean±SEM. Sample numbers are indicated in *n*. The velocity distributions (left panels in **A** and **B**) were well characterized by Gaussian distributions (solid lines). At a given ATP concentration, the average bead velocity did not differ appreciably for two- *vs.* single- kinesin transport. For single-kinesin transport (right panels in **A**), at all three ATP concentrations, the measured travel distributions were well described by a single exponential decay (solid lines). The average travel distance of beads driven by a single kinesin remained constant over the ATP concentration examined (1.67μm). For two-kinesin transport (right panels in **B**), travel distribution at 1mM ATP was well characterized by a single exponential decay (solid line, see text and Fig. 5 for detailed discussion). At 10μM and 20μM ATP, the number of two-motor beads traveling out of camera view (~7.6μm) became significant, and the total corresponding events are indicated in a hatched bar at 8μm. We observe a strong, negative correlation between velocity and two-kinesin travel.

**Figure 5**. Quantitative comparison between theory (Eqn. 4, (*7*)) and measurements (Fig. 4B). (**A**) Least-$\chi^2$ fitting of Eqn. (4) to the measured two-motor travel distribution at 1mM ATP constrained individual motor's binding rate $\pi_{ad}$ to 0.71/s in the two-motor ensemble. (**B**) Predicted decay lengths $\lambda_a$ and $\lambda_b$ as a function of motor velocity, using the experimentally constrained $\pi_{ad}$=0.71/s. The two decay lengths at $v_{max} = 0.9$μm/s are indicated in scatter. (**C**) Predicted two motor travel distributions for the three velocities examined in the current study. Within our measurement window (0.6-7.6μm, grey region), the predicted distributions at all three velocities measured are dominated by the single exponential with decay length $\lambda_a$. Measured travel distribution of two-kinesin transport at $v_{max} = 0.9$μm/s is shown as scatter. (**D**) Predicted *vs.* measured fraction of long travels at $v =$0.26μm/s and at $v =$0.16μm/s. The fraction of long travels is defined as the fraction of travel exceeding 7.6μm in all travels measured (greater than 0.6μm). Measurement error was determined as $\sqrt{p(1-p)/n}$, where *p* is the measured long travel fraction, and *n* the measurement sample size (see Fig. 4B). The predicted populations were calculated using

the predicted two-motor travel distributions in (**C**), and are in excellent agreement with measurements.

**SUPPLEMENTAL INFORMATION**

**Figure S1**. Simulated *vs*. measured single-motor travel distance for kinesin. Shown are mean±SEM, and simulation sample size. To evaluate single motor travel distances, we employed a previously developed Monte Carlo algorithm detailed in (*10*). We utilized the same parameters as described for kinesin in (*10*) with the following exceptions. We constrained the unbinding rate $\varepsilon$ at $v_{max} = 0.9$μm/s, using single-kinesin travel measured at 1mM ATP. We then allowed the unbinding rate to either vary linearly with velocity (magenta bars), or to remain constant regardless the motor's speed (cyan bars). Our simulation demonstrates explicitly that a velocity-independent unbinding rate does not agree with the measured single-kinesin travel distance (grey bars). Intuitively, the time required to cover a particular distance is linearly proportional to velocity. Thus, with a velocity-independent unbinding rate, the single-motor travel distance should reduce linearly with reducing velocity. This does not agree with measurements (cyan *vs*. blue bars). A velocity dependent unbinding rate ($\varepsilon = v/d$), on the other hand, recaptures the experimental measurements quite well (magenta *vs*. grey bars).

# Figure 1

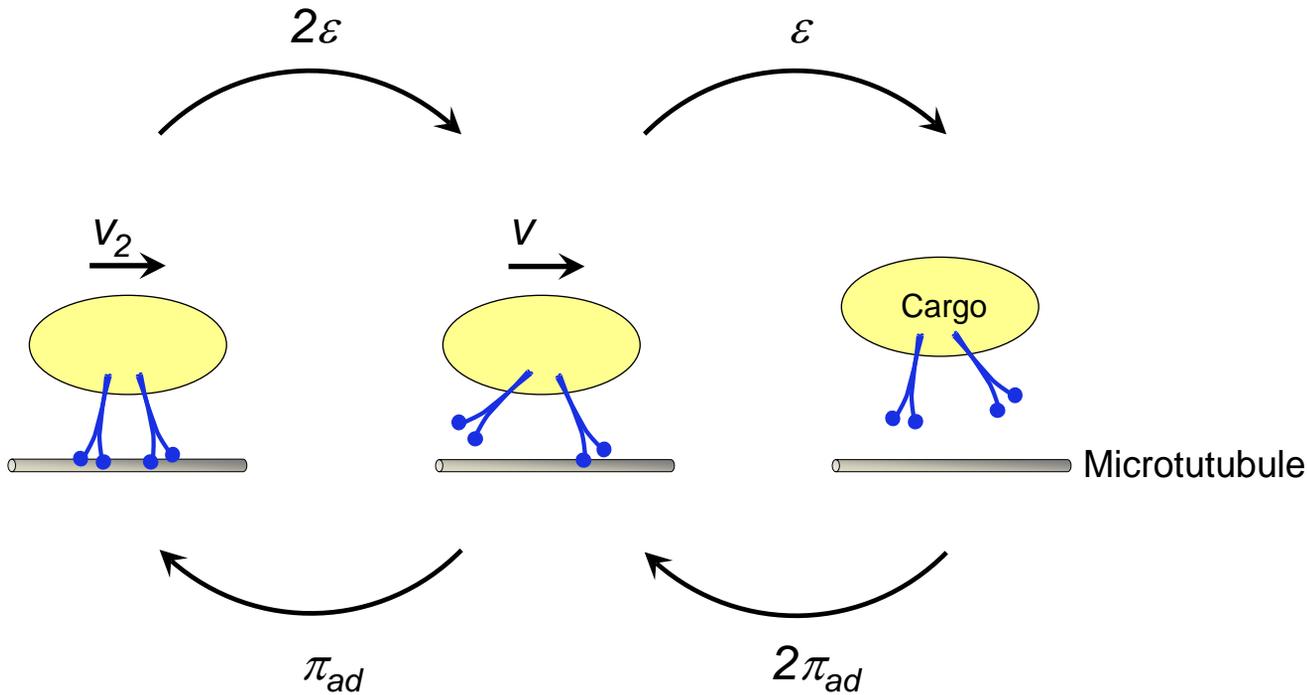

$\varepsilon$ : single-motor unbinding rate (1/s)
$\pi_{ad}$ : single-motor binding rate (1/s)
$v$ : single-motor velocity (μm/s)
$v_2$ : cargo velocity when both motors are attached (μm/s)

# Figure 2

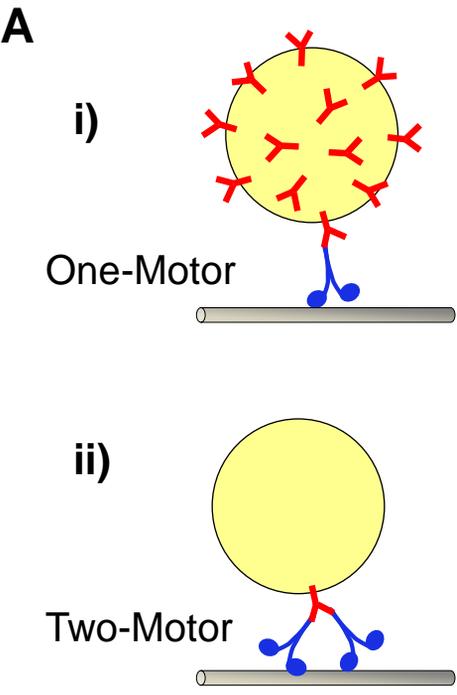
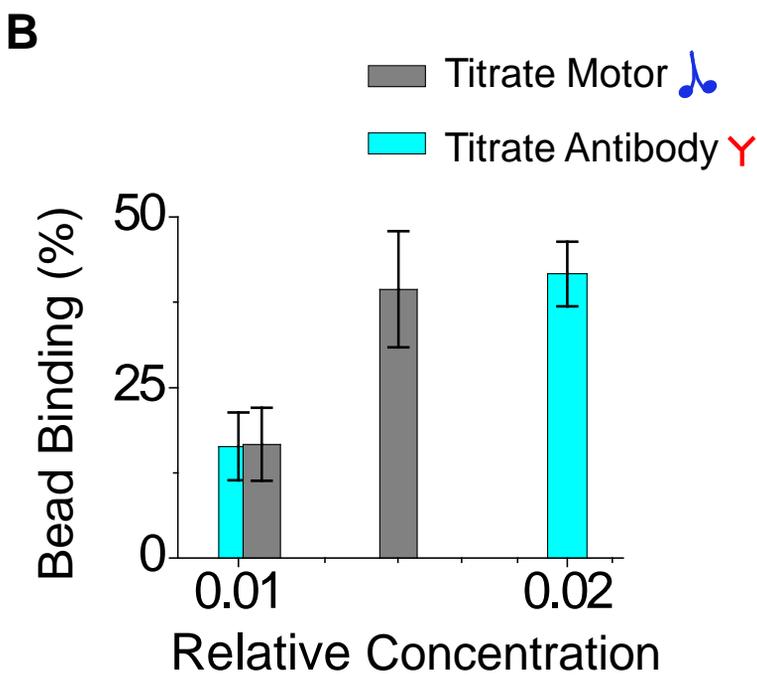

# Figure 3

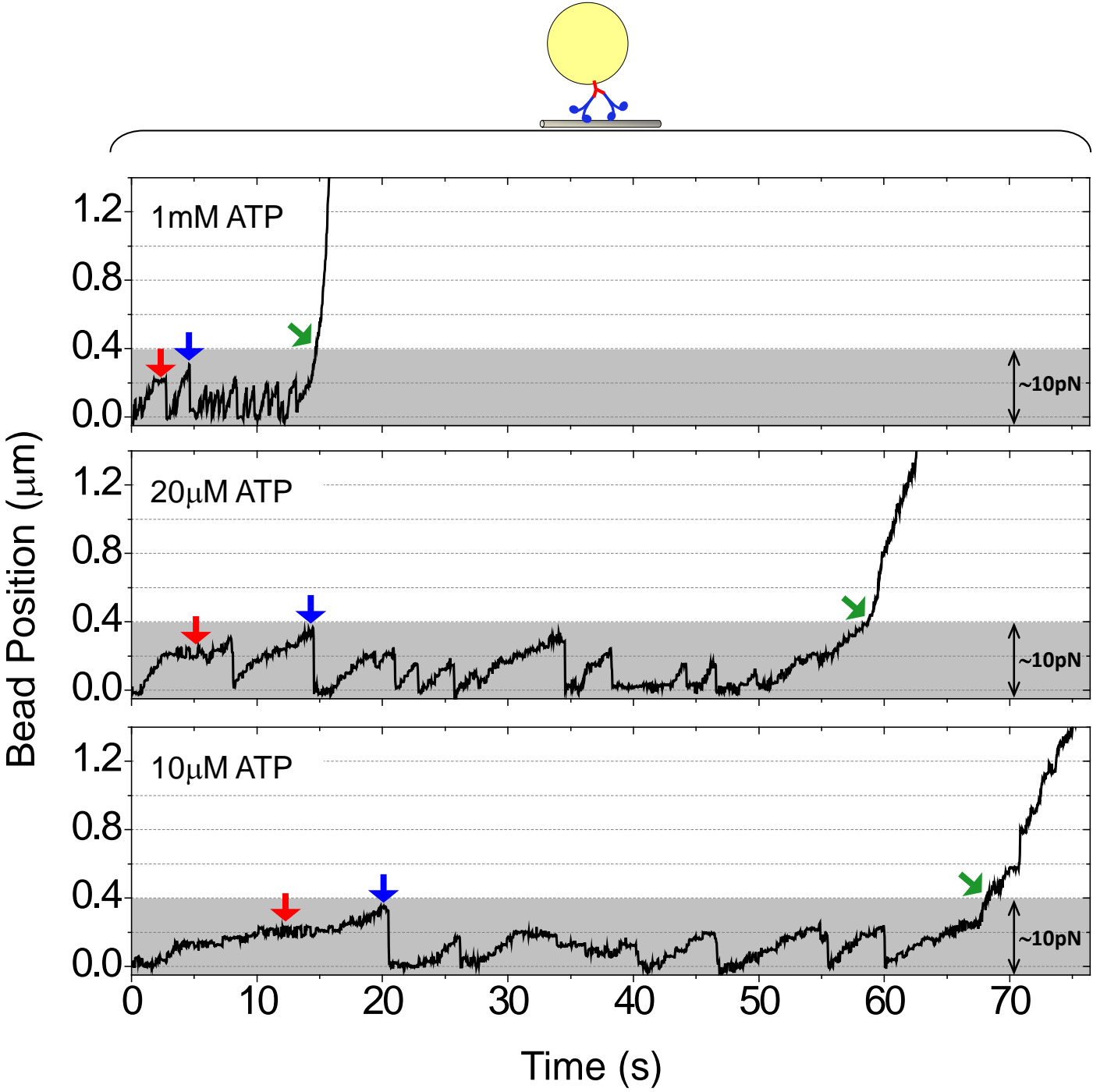

# Figure 4

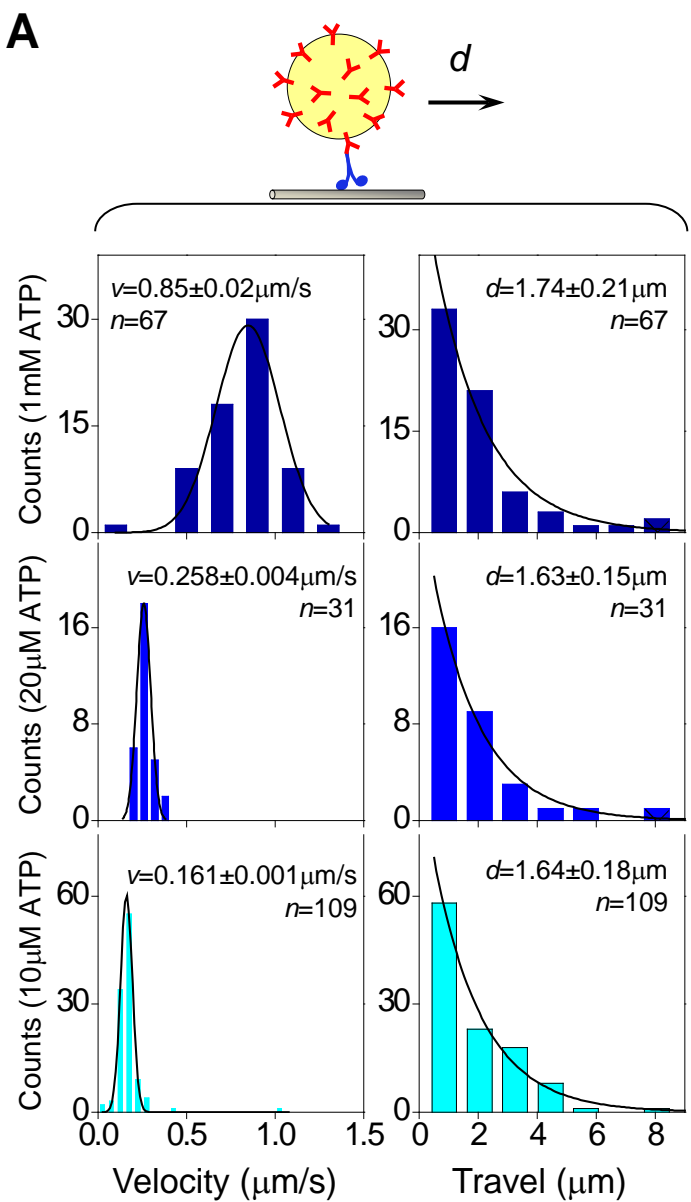
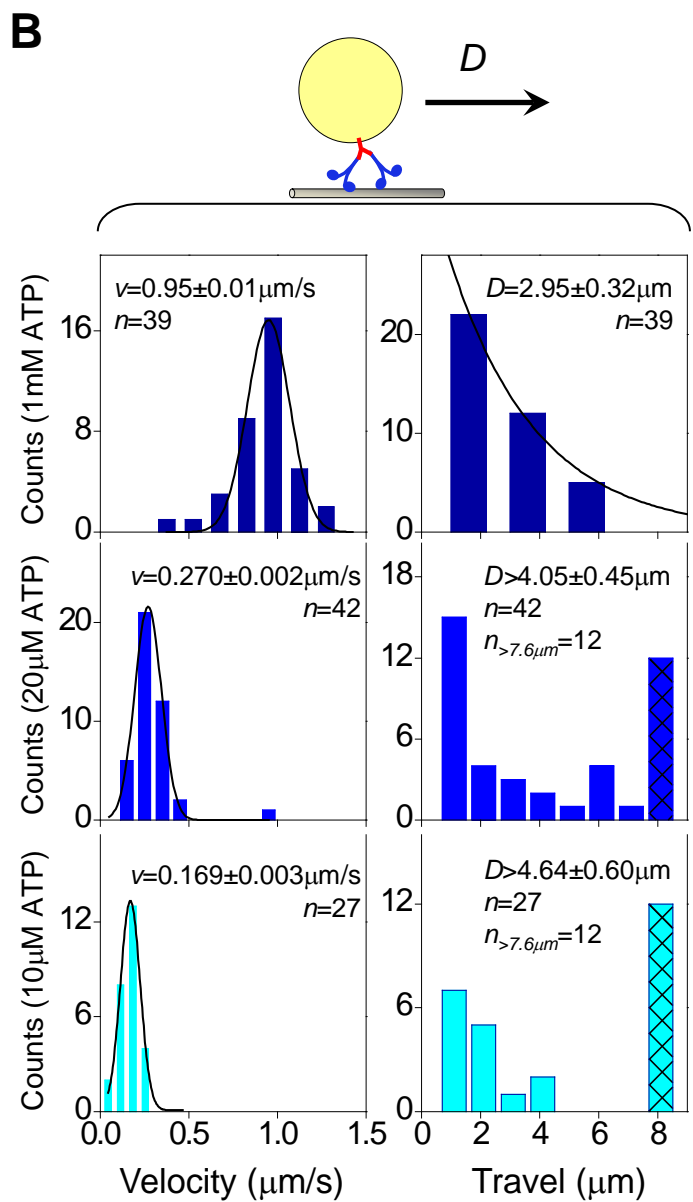

# Figure 5

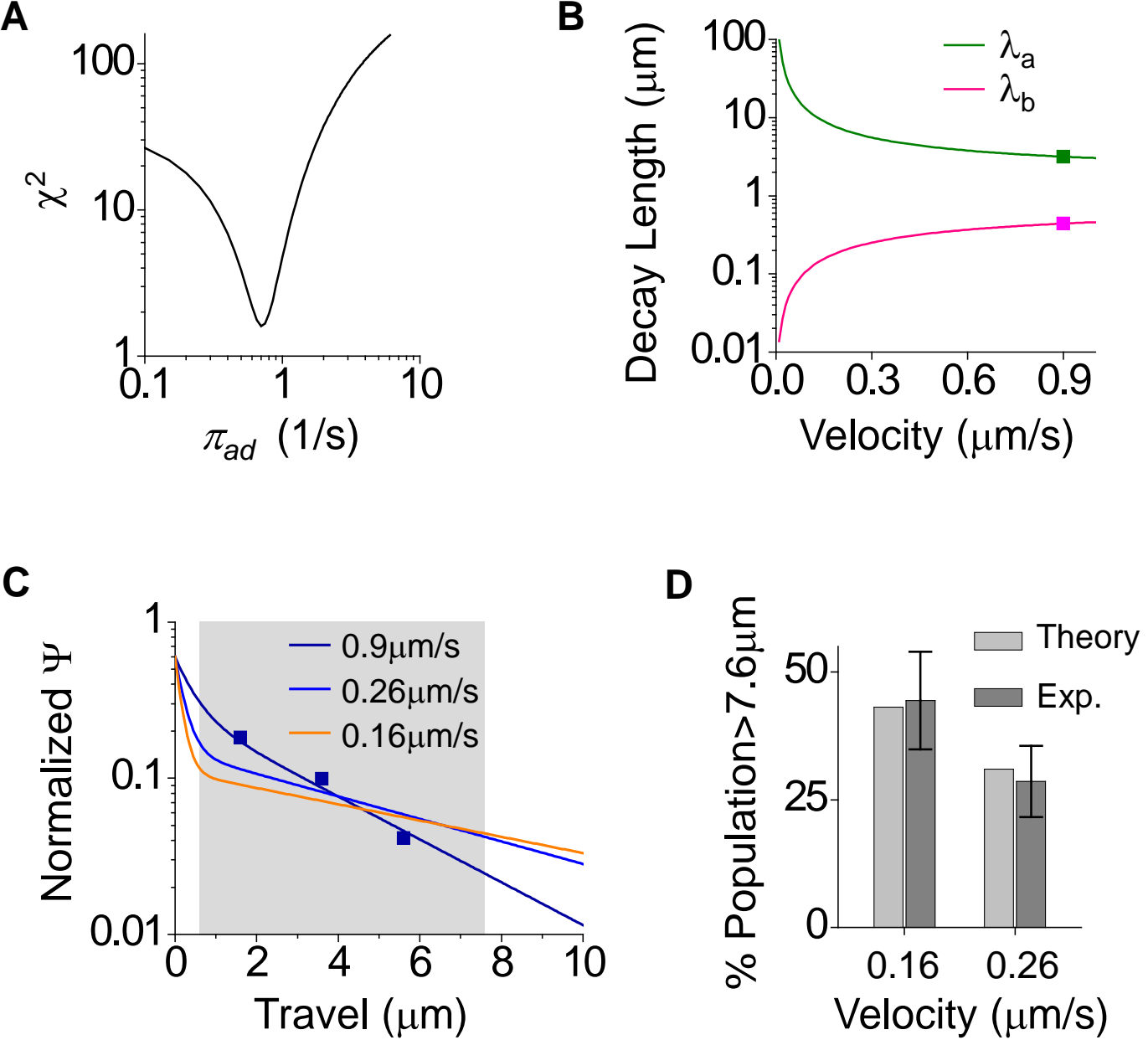

# Supplemental Figure 1

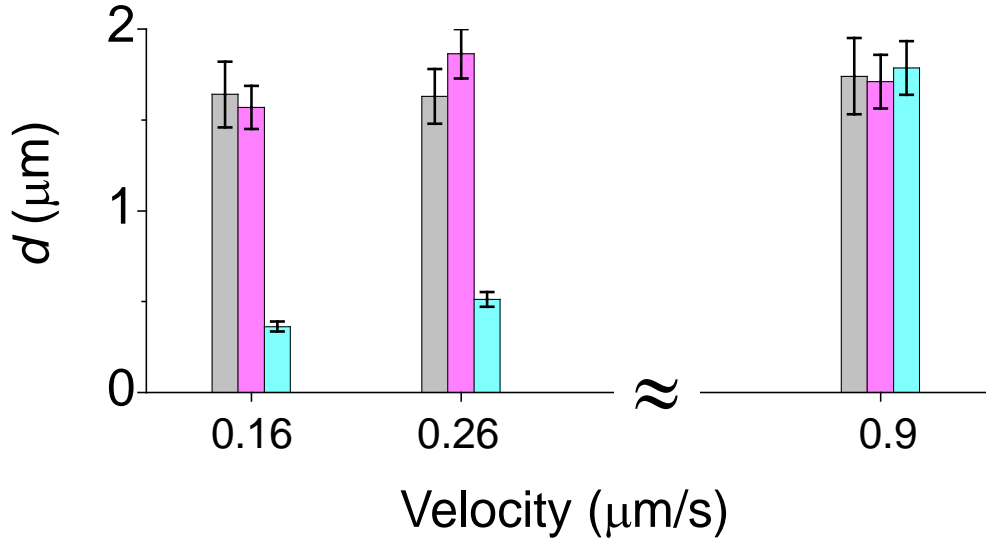